# Geminate recombination of electrons generated by above-the-gap (12.4 eV) photoionization of liquid water. [1]

Rui Lian, Dmitri A. Oulianov, Ilya A. Shkrob, and Robert A. Crowell[*]

*Chemistry Division, Argonne National Laboratory, Argonne, IL 60439*

**Abstract**

The picosecond geminate recombination kinetics for hydrated electrons generated by 200 nm two photon absorption (12.4 eV total energy) has been measured in both light and heavy water. The geminate kinetics are observed to be almost identical in both $H_2O$ and $D_2O$. Kinetic analysis based upon the independent reaction time approximation indicates that the average separation between the electron and its geminate partners in $D_2O$ is 13% narrower than in $H_2O$ (2.1 nm vs. 2.4 nm). These observations suggest that, even at this high ionization energy, autoionization of water competes with direct ionization.

___________________________________________________________________

[*] To whom correspondence should be addressed: *Tel* 630-252-8089, *FAX* 630-2524993, *e-mail:* rob_crowell@anl.gov.

## 1. Introduction.

Although recombination dynamics of photoelectrons in liquid water have been extensively studied (e.g., refs. [1-8]), many questions concerning the mechanism of generation, localization, and solvation of these electrons still remain unanswered. It is presently accepted that the time profile of geminate recombination kinetics for hydrated electrons generated by multiphoton laser ionization of liquid water, reaction (1),

$$H_2O_{liq} \xrightarrow{n\ h\nu} e^-_{aq} + H_3O^+ + OH^\bullet \qquad (1)$$

depends on the total photoexcitation energy $E_{tot}$. [1-5] Two- or three-photon excitation of water at lower photon energy (so that $E_{tot} < 9.2$ eV) produces electrons that are, on average, just 0.9-1 nm away from their parent hole, [1-6] whereas at higher total photon energy ($E_{tot} > 11$ eV), the photoelectrons are ejected at least 2 nm away. [1,2,3,7,8] It is believed that in this high-energy regime, the electrons might be ejected directly into the conduction band of the solvent, [1,3] as the conduction band of water is commonly placed between 9 and 10 eV above the ground state. [9,10] At lower energy, the photoionization involves concerted proton and electron transfers, [9-12] perhaps to pre-existing traps. [13,14] In the intermediate regime, autoionization of water is thought to compete with these two photoprocesses. [3]

In this work we present picosecond geminate recombination kinetics for electrons in light and heavy water obtained using a total excitation energy of 12.4 eV, which is well above the accepted band gap of liquid water. The 12.4 eV excitation energy was attained via two photon absorption using a subpicoseond 6.2 eV (200 nm) light source. Under these photoexcitation conditions, direct ionization of the solvent (by which we mean the ejection of photoelectron directly into the conduction band of the liquid) has been thought to prevail. [3] Due to the smaller vibration energy of accepting O-D modes in heavy water, and, therefore, less efficient inelastic scattering of the extended-state (quasifree) electron, it has been expected that the thermalization of the conduction band electrons in D$_2$O (generated by direct photoionization of the solvent) would be less efficient, and the average thermalization path in D$_2$O would be *longer* than in H$_2$O. [3,15] Pulse radiolysis studies have shown that at high excitation energy the distribution of thermalization distances is broad (2-3 nm width in light water) and it is indeed ca. 30% *broader* for heavy water. [15] However, this result does not indicate under what conditions it becomes possible to produce such electron distributions in the course of *photo-ionization*. The required energy might be so high that more than one electron is generated per excitation event (the average energy needed for water ionization in radiolytic spurs is 20-25 eV [15]). Still, it is generally believed that for higher excitation energy, the electron distributions generated via photoionization of water would more closely follow the electron distributions obtained via radiolysis. As shown below, for the case of 12.4 eV photoionization these expectations are not supported by our observations, indicating that even higher excitation energies are needed to observe the same results as those obtained by pulse radiolysis. Whereas above-the-gap excitation of water produces broad electron distributions that are similar to those observed in radiolytic spurs, the distribution is, actually, *narrower* for heavy water.

**2. Experimental.**

Deionized water with conductivity < 2 nS/cm was used in all experiments with $H_2O$. An $N_2$-saturated 1 L sample was circulated using a gear pump through a jet nozzle. A 500 mL sample of heavy water (99 atom %, Aldrich) was used in all experiments with $D_2O$. No change in the kinetics was observed after continuous photolysis of this sample. The details of the flow system are given elsewhere. [16]

The picosecond transient absorption (TA) measurements were carried out using a 1 kHz Ti:sapphire setup, details of which are given in refs. [16,17,18]. This setup provided 60 fs FWHM, 3 mJ light pulses centered at 800 nm. One part of the beam was used to generate probe pulses while the other part was used to generate the 200 nm (fourth harmonic) pump pulses. Up to 20 µJ of the 200 nm light was produced in this way (300-350 fs FWHM pulse). The pump and probe beams were perpendicularly polarized and overlapped at the surface of a 90 µm thick high-speed water jet at $5^o$. No change of the TA kinetics with the polarization of the probe light was observed. The pump power, before and after the sample, was monitored using a thermopile power meter. The TA traces were obtained using a 1-5 µJ, 200 nm pump pulse focused, using a thin $MgF_2$ lens, to a round spot of 300 μm FWHM; the probe beam was typically 50-60 μm FWHM. Radial profiles of these beams at the jet surface were obtained by scanning a 10 μm pinhole across the beam. A typical TA signal ($\Delta OD_\lambda$, where $\lambda$ is the wavelength of the probe light in nanometers) at the maximum was 10-to-60 mOD. The vertical bars in the figures represent 95% confidence limits for each data point. 150-200 delay time points acquired on a quasi-logarithmic grid were used to obtain the decay kinetics of the electron out to 500 ps.

## 3. Results.

Fig. 1(a) shows the 200 nm pulse energy dependence for the normalized TA signal $\Delta OD_{800}(t)$ obtained at the delay time $t$=10 ps (this dependence is given on a double logarithmic plot). At this delay time, the thermalization is complete [19] (in accord with the previous observations for photoionization at lower excitation energies, see refs. [4,6,8,11,20,21]) and the TA signal is from fully hydrated electron, $e_{aq}^-$. The initial slope of this plot (solid line in Fig. 1(a)) is close to 1.84±0.07, indicating *biphotonic* ionization of water by the 200 nm light. At high power, this slope decreases to 1-1.5 due to nonuniform absorbance of the 200 nm light by the sample. [23] When $\Delta OD_{800}(t = 10 \ ps) > 0.5$, the sub-nanosecond kinetics plotted on the logarithmic time scale show the characteristic bend down typical of bimolecular cross recombination (i.e., recombination that occurs between two different ionization events) in the sample bulk (compare traces (i) and (ii) in Fig. 1(b)). [22,23] These TA kinetics are power-dependent and their time profiles are different from the profile of *geminate* recombination kinetics obtained in the low-power regime. In the latter regime, normalized TA kinetics are independent of the 200 nm pulse energy. Only the pump power independent kinetics are considered in the rest of this paper. The concentration of $e_{aq}^-$ produced in the experiment was on the order of $10^{-5}$ mol/dm$^3$.

Typical geminate recombination kinetics of $e_{aq}^-$ on a sub-nanosecond time scale (for $\lambda$=0.8 μm) are shown in Fig. 2. Very similar kinetics for $t$>5 ps were observed at other probe wavelengths, including $\lambda$=266 nm at which both $e_{aq}^-$ and OH radical absorb (the 266 nm light was obtained by third harmonic generation). [16,17,18] The comparison between the normalized recombination kinetics of $e_{aq}^-$ in $H_2O$ and $D_2O$ shown in Fig. 2 suggests these two traces are identical within the confidence limits of our measurement. Note that the origin of the vertical axis corresponds to 85% survival fraction.

Using the "independent reaction time" (IRT) model developed by Pimblott, [24] which is commonly used to simulate the geminate dynamics of water spurs, the decay kinetics of $e_{aq}^-$ for a trial distribution $4\pi r^2 P(r)$ of the initial thermalization distances of the electron can be obtained. Typically, a Gaussian distribution, $P(r) \propto \exp(-r^2/2\sigma_G^2)$, has been used for such simulations. [1,2,24] Since the diffusion coefficients for the electron and rate constants for recombination reactions

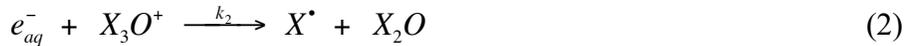

$$e_{aq}^- + X_3O^+ \xrightarrow{k_2} X^\bullet + X_2O \qquad (2)$$

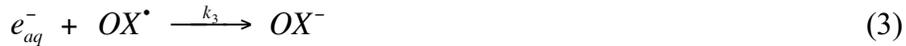

$$e_{aq}^- + OX^\bullet \xrightarrow{k_3} OX^- \qquad (3)$$

(where X = H or D) for light and heavy water are different (Table 1), different recombination kinetics are expected to be observed in $H_2O$ and $D_2O$ for the same initial distribution $P(r)$. Conversely, identical recombination kinetics imply different initial distributions $P(r)$. Fitting the kinetics by the IRT model equations (see refs. [1,2,24] and caption to Fig. 1 and Table 1 for the parameters and details of the simulation) yields $\sigma_G$=2.4 nm for $H_2O$ and $\sigma_G$=2.1 nm for $D_2O$. Thus, the electron distribution in the heavy water is narrower than in the light water. While the quality of these kinetic traces does not allow us to put accurate confidence limits on these estimates, it is clear from Fig. 1 that only curves with $\sigma_G(D_2O) < \sigma_G(H_2O)$ can account for these data. As will be discussed below, this result is surprising since under our photoexcitation conditions (12.4 eV total excitation energy), direct ionization of the solvent is thought to dominate. [3,5,20]

**4. Discussion.**

Recently, Sander et al. [3] have analyzed the existing data on the H/D isotope effect $\alpha_{H/D} = \Omega_r(H_2O)/\Omega_r(D_2O) - 1$ for the probability $\Omega_r$ of electron recombination at $t = \infty$ as a function of the total excitation energy. The probability of electron recombination, $\Omega_r$ is determined by the width of the initial electron distribution, $\sigma_G$. Sander et al. [3] suggested that the switchover from positive to negative $\alpha_{H/D}$ with increasing energy (which is observed between 9.5 and 10 eV) reflects a competition between the photoinduced electron transfer to a pre-existing trap (that prevails for $E_{tot} <$ 9.5 eV) and autoionization of water (that prevails for $E_{tot} > 9.5$ eV). The latter is initiated by a *bound-to-bound* transition to a short-lived excited state; this excited state promptly dissociates yielding a conduction band electron. Sander et al. [3] predicted the second change of the sign of $\alpha_{H/D}$, from negative to positive, when $E_{tot} > 11$ eV and direct ionization (which involves a *bound-to-free* transition) prevails.

At first glance, our 12.4 eV data lend support to the suggestions made by Sander et al. [3]: $\alpha_{H/D} \approx 0$ (in agreement with the predicted second switchover) and the widths $\sigma_G$ compare well with those in the radiolytic spurs (suggesting the onset of direct photoionization). On the other hand, even at this high excitation energy the distribution of electrons in D$_2$O is 13% narrower than in H$_2$O; this difference is comparable to 15% obtained by Sander et al. for 10 eV photoexcitation [3] (where autoionization is supposed to prevail). The results presented here do no support the idea that the energy at which the thermalization distance becomes greater in D$_2$O than in H$_2$O occurs at ca. 11 eV, in fact it does not even occur at 12.4 eV. The turn over in the isotope effect is believed to be evidence for the conversion of the ionization mechanism from autoionization to primarily direct ionization. Either direct ionization is not dominant at 12.4 eV or it does not result in a longer thermalization path for D$_2$O. Both of these possibilities hint at a complex picture of water ionization at high excitation energy.

There have been other results suggesting such a complexity. Synchrotron radiation studies of Brocklehurst [25] reveal that that the delayed luminescence from recombination of geminate ion pairs generated by vacuum UV photoexcitation of a viscous hydrocarbon squalane continuously change between 10 and 25 eV, well above

6.

the gap energy of this liquid (8-9 eV). The observed change in the recombination dynamics indicates continuous broadening of the electron distribution with increasing energy. No such broadening is observed for UV and vacuum UV photoionization of aromatic solutes in liquid alkanes; the width of the distribution stabilizes at 3 eV above the ionization threshold. [25] These results suggest that in squalane, direct ionization competes with some other photoionization mechanism, even at these high photon energies. [12,25,26] The magnetic field effect data of Brocklehurst [25] and Jung and coworkers [26] provide evidence for the existence of an ionization channel through which the singlet correlation between the radical ions in the geminate pair is rapidly lost; spin-orbit interaction in the excited bound state of the solvent (which mediates the autoionization) has been suggested as such a channel. [25] This loss of spin correlation is quite notable between 11 and 16 eV, at the very onset of the direct ionization. [25] Perhaps, the distinction between direct ionization and autoionization is not as clear-cut as suggested by Sander et al.: [3] both photoprocess compete well above the postulated 11 eV threshold.

**5. Conclusion.**

The geminate recombination kinetics for hydrated electrons generated by absorption of two 200 nm quanta (12.4 eV total energy) by light and heavy water are similar. Kinetic analyses within the framework of the IRT model [24] suggest that the initial distribution of distances between the electron and its geminate partners in $D_2O$ is narrower than in $H_2O$. Since the opposite trend is expected in the regime where direct ionization prevails, it appears that autoionization of water still competes with direct ionization at this high excitation energy. The latter ionization mechanism has been thought to prevail when the total excitation energy is greater than 11 eV. [1,3,5,20] While our result is unexpected, it is consistent with the previous observations of competing ionization channels in the vacuum UV photoexcitation of molecular liquids. [25,26]

**6. Acknowledgement.**

We thank Drs. C. D. Jonah, D. M. Bartels, and S. Pommeret for valuable discussions. The research was supported by the Office of Science, Division of Chemical Sciences, US-DOE under contract number W-31-109-ENG-38.

**Figure captions.**

**Fig. 1**

(a) The dependence of the TA signal ($\lambda$=800 nm) from hydrated electron in liquid H$_2$O at 25 °C (observed at $t$=10 ps after the photoionization) as a function of the power of the 200 nm pulse *(open circles)*. For this measurement, the pump and probe beams were tightly focussed. Note the double logarithmic scale. The initial slope of 1.84±0.07 is consistent with biphotonic ionization of water by 200 nm light. (b) Normalized TA kinetics at the extremes of the dynamic range are shown in (a). The traces were normalized at $t$=5 ps. Trace (i) *(open squares)* was obtained for the highest pump power; trace (ii) *(open circles)* was obtained for the lowest pump power in (a). The solid lines are guides to the eye. The vertical bars are 95% confidence limits. Note the logarithmic time scale and non-zero origin of the vertical axis. The rapid decay on the subnanosecond time scale is due to the cross-recombination in the water bulk.

**Fig. 2.**

Geminate recombination dynamics of hydrated electrons generated by bi- 200 nm photon excitation of neat liquid H$_2$O and D$_2$O at 25 °C. The photoinduced optical density $\Delta OD_{800}$ (observed at $\lambda$ = 800 nm and normalized at $t$=10 ps) is plotted vs. the delay time $t$ on a logarithmic time scale out to 500 ps. Note that the origin of the vertical axis is at 85% recombination efficiency. The data for heavy water are indicated using filled circles and diamonds, for light water - open squares and triangles. The vertical bars indicate 95% confidence limits. Two series of data obtained at 2 µJ and 5 µJ excitation power (*circles and squares* and *diamonds and triangles*, respectively) are shown to illustrate the constancy of the time profile with the 200 nm light radiance in the explored excitation regime. All four normalized kinetics are identical within the confidence limits. Solid lines are IRT model simulations. The simulation parameters are given in Table 1 in the supplement. The black solid line is the simulation for light water with $\sigma_G$ = 2.4 nm (average separation of 3.83 nm); traces (i), (ii), and (ii) are simulations for heavy water for $\sigma_G$ = 2.1, 2.4, and 2.7 nm, respectively.

**Table 1.**

Simulation parameters for IRT modeling of recombination dynamics (X=H or D) of hydrated electron in room-temperature liquid water (after refs. [1,2,24])

| parameter | $H_2O$ | $D_2O$ |
|---|---|---|
| diffusion coefficients, $\times 10^{-5}$ cm$^2$/s | | |
| hydrated electron | 4.9 | 3.9 |
| hydronium ion | 9.0 | 6.7 |
| hydroxyl radical | 2.8 | 2.2 |
| reaction constants, etc. [a] | | |
| $k_2$, $\times 10^{10}$ M$^{-1}$ s$^{-1}$ | 2.3 | 1.0 |
| $k_3$, $\times 10^{10}$ M$^{-1}$ s$^{-1}$ | 3.1 | 2.5 |
| reaction velocity $v$ (rxn. (2)), m/s | 4.0 | 1.5 |
| other simulation parameters [b] | | |
| $\sigma_G$, nm [c] | 2.4 | 2.1 |

(a) Reaction radii of 0.5 and 0.54 nm were assumed for reactions (2) and (3), respectively.

(b) Dielectric constant of 78 and initial $X_3O^+...OX$ distance of 0.28 nm were assumed for both liquids.

(c) The width of the $r^2$-Gaussian electron distribution.

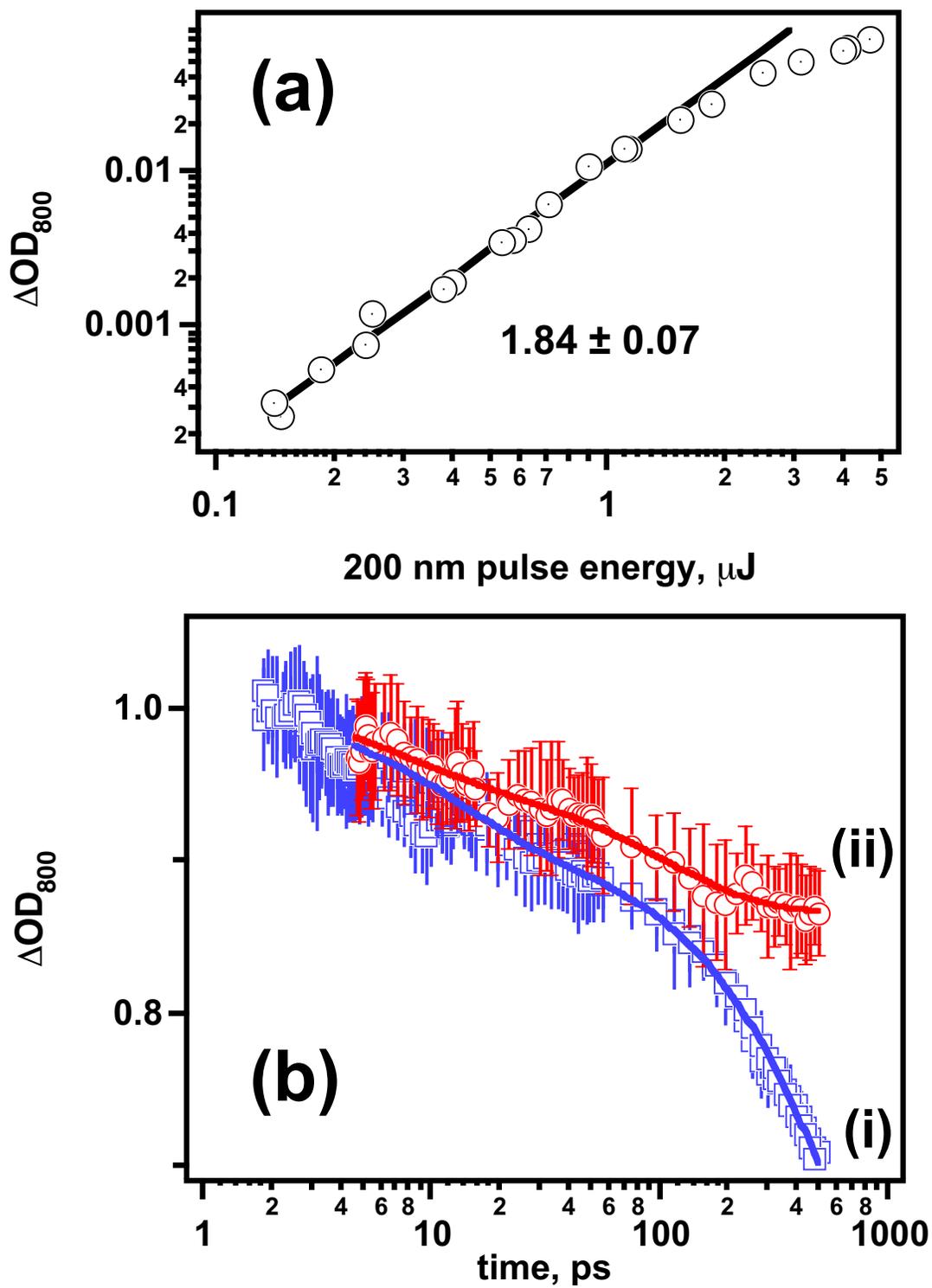

**Figure 1; Lian et al.**

13.

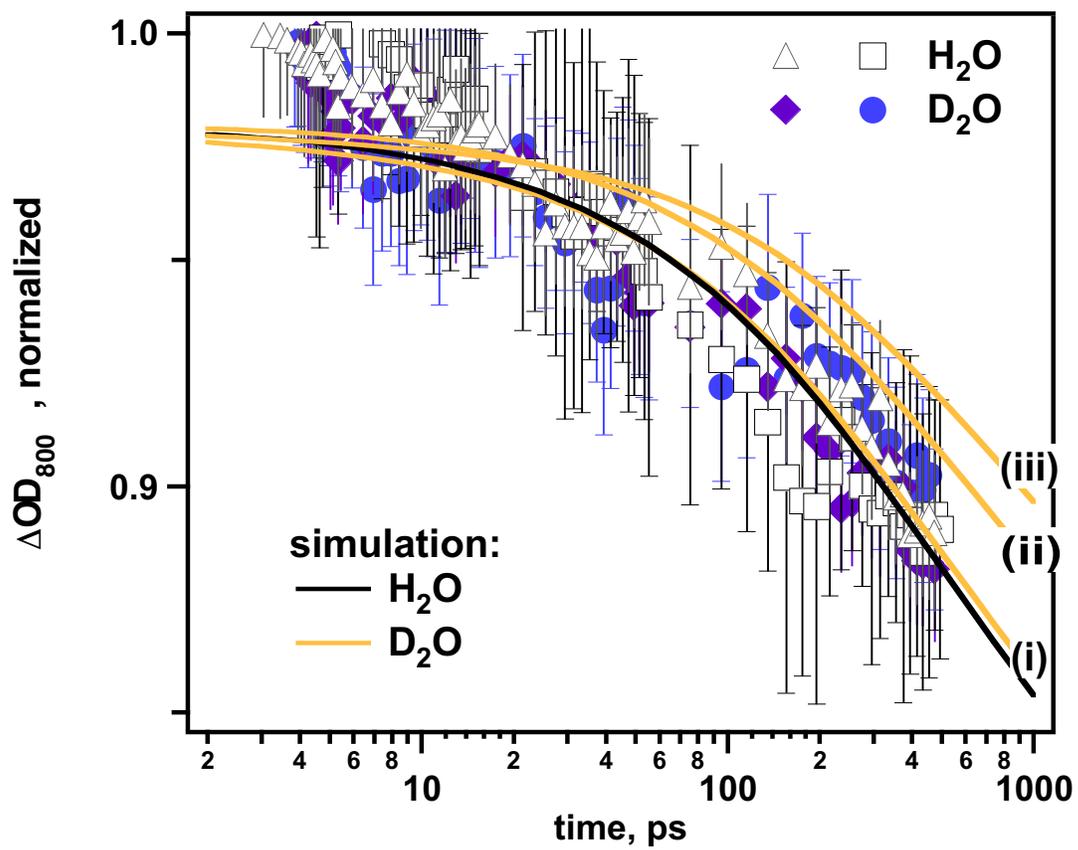

**Figure 2; Lian et al.**